\documentstyle[12pt]{article}

\begin{document}
\newcommand{\beq}{\begin{equation}}
\newcommand{\eeq}{\end{equation}}
\newcommand{\al}{\alpha}
\newcommand{\be}{\beta}
\newcommand{\de}{\delta}
\newcommand{\s}{\sigma}
\newcommand{\r}{\rho}
\newcommand{\om}{\omega}
\newcommand{\e}{\eta}
\newcommand{\th}{\theta}
\newcommand{\g}{\gamma}
\newcommand{\Tr}{{\rm Tr}\,}
\begin{flushright}
March 1998\\
%(NBI-HE-97-06)
\end{flushright}
\vspace{0.5cm}
\begin{center}
\large
THERMODYNAMIC PROPERTIES OF THE\\ PIECEWISE UNIFORM STRING\\[0.5cm]

I. Brevik\footnote{E-mail address: Iver.H.Brevik@mtf.ntnu.no}\\
Applied Mechanics, Norwegian University of Science\\
and Technology, N-7034 Trondheim, Norway

A.A. Bytsenko\footnote{E-mail address: abyts@fisica.uel.br
\,\,\,\,\, On leave from Sankt-Petersburg State Technical University}\\
Departamento de Fisica, Universidade Estadual de Londrina,\\
Caixa Postal 6001, Londrina-Parana, Brazil

and\\
H.B. Nielsen\footnote{E-mail address: hbech@nbi.dk}\\
The Niels Bohr Institute, Blegdamsvej 17, DK-2100\\
Copenhagen, Denmark\\[0.5cm]
\end{center}
\begin{center}
\end{center}

\begin{abstract}

The thermodynamic free energy $F$ is calculated for a gas whose particles
are the quantum excitations of a piecewise uniform bosonic string. The
string consists of two parts of length $L_I$ and $L_{II}$, endowed with
different tensions and mass densities, adjusted in such a way that the
velocity of sound always equals the velocity of light. The explicit
calculation is
done under the restrictive condition that the tension ratio
$x = T_I/T_{II}$ approaches zero. Also, the length ratio $s = L_{II}/L_I$ is
assumed to be an integer. The expression for $F$ is given on an integral form,
in which $s$ is present as a parameter. For large values of $s$, the 
Hagedorn temperature becomes proportional to the square root of $s$.

\end{abstract}
\newpage

\section{Introduction}

In conventional theories of the bosonic string in D-dimensional spacetime the
string is taken to be {\it uniform}, throughout the whole of its length $L$.
The {\it composite} string model, in which the (relativistic) string is
assumed to consist of two or more separately uniform pieces, is a variant
of the conventional theory. The composite model was introduced by the
present authors in 1990 \cite{brevik90};
the string was there assumed to consist of
two pieces $L_I$ and $L_{II}$. The dispersion equation was derived, and
the Casimir energy calculated for various integer values of the length ratio
$s = L_{II}/L_I$. Later on, the composite string model has been generalized
and further studied from various points of view \cite{li91} -
\cite{berntsen97}.

It may be useful to summarize some reasons why the composite string model
turns out to be an attractive model. First, if one does Casimir energy
calculations one finds that the system is remarkable easy to regularize:
one has access to the cutoff method \cite{brevik90}, the complex contour
integration method \cite{brevik94} - \cite{brevik96}, \cite{brevik},
or the Hurwitz $\zeta$-function method \cite{li91}, \cite{brevik95},
\cite{brevik96}, \cite{brevik}. (Ref. \cite{berntsen97} contains a review of
the various regularization methods.)
As a physical result of the Casimir energy calculations it is also worth
noticing that the energy is in general nonpositive, and is more negative the
larger is the number of uniform pieces in the string. Second, the composite
string model may serve as a useful two-dimensional field theoretical model of
the early universe. These aspects have recently been discussed more closely
in \cite{bayin96}. Finally, as a possible practical application of the
composite
string model, we mention the recent attempt \cite{elizalde96}
that has been done
to analyse the influence from the Casimir effect for uniform strings on
the swimming of micro-organisms. Probably would here a composite string model
be more adequate than the very simple uniform string model.

The purpose of the present paper is to calculate thermodynamic quantities -
the free energy (the one-loop partition function) for the composite 
two-piece string. 
It ought to be emphasized that the present string model is relativistic, in
the sense that the velocity of sound is always taken to be equal to $c$.
As far as we are aware, this task has not been undertaken before. When we come
to concrete calculations, we shall put $D = 26$. Moreover we shall limit
ourselves to the limiting case in which the tension ratio, defined as
$x = T_I/T_{II}$, goes to zero. As shown earlier \cite{brevik90}, this
case leads to significant simplifications in the formalism, but is yet
nontrivial enough to show the essential physical behaviour of the system.
Another simplification is that we shall assume $s$ to be an integer.

In the next section we discuss briefly the essentials of the theory of the
classical planar string: the general dispersion equation, and the junction
conditions as well as the eigenvalue spectrum in the case of $x \rightarrow
0$. In Sec. \ref{III} we develop the classical theory of the composite string
in flat $D$-dimensional spacetime. The string coordinates $X^{\mu}$ are
expanded into oscillator coordinates, and formulas are derived for the
Hamiltonian and the string mass.
In Sec. \ref{IV} we derive the quantum theory, assuming $D = 26$, focusing
attention on the free energy $F$ and the critical Hagedorn temperature.
Finally we end with some conclusions in Sec. 5.

\section{Planar Oscillations of the Classical String in the Minkowski
Space}\label{II}

We begin by considering the classical theory of the oscillating two-piece
string in the Minkowski space. The total length of the string is $L$.
For later purpose we shall set  $L = \pi$.  With $L_I$, $L_{II}$ denoting the length of the two pieces, we
thus have $L_I + L_{II} = \pi$. As mentioned the string is relativistic,
in the sense that
the velocity $v_s$ of transverse sound is everywhere required to be equal
to the velocity of light ($\hbar = c = 1$):
\beq
v_s = (T_I/\r_I)^{1/2} = (T_{II}/\r_{II})^{1/2} = 1.
\label{1}
\eeq
Here $T_I, T_{II}$ are the tensions and $\r_I, \r_{II}$ are the mass
densities of the two pieces. We let $s$ denote the length ratio and $x$ the
tension ratio:
\beq
s = L_{II}/L_I, ~~~~ x = T_I/T_{II}.
\label{2}
\eeq
Assume now that the transverse oscillations of the string, called
$\psi(\s,\tau)$, are linear, and take place in the plane of the string.
(We employ usual notation, so that $\s$ is the position coordinate and
$\tau$ the time coordinate of the string.) We can thus write in the two
regions
\begin{eqnarray}
\psi_I & = & \xi_I e^{i\om(\s-\tau)} + \e_I e^{-i\om(\s+\tau)},\cr
\psi_{II} & = & \xi_{II} e^{i\om(\s-\tau)} + \e_{II} e^{-i\om(\s+\tau)},
\label{3}
\end{eqnarray}
with the $\xi$ and $\e$ being constants. Taking into account the junction
conditions at $\s = 0$ and $\s = L_I$, meaning that $\psi$ itself as well
as the transverse force $T \partial \psi / \partial \s$ be continuous, we
obtain the dispersion equation
\beq
\frac{4x}{(1-x)^2} \sin^2\frac{\om\pi}{2} +
\sin \left( \frac{\om\pi}{1+s} \right)
\sin \left( \frac{\om s \pi}{1+s} \right) = 0
\label{4}
\eeq
(more details are given in \cite{brevik90}). From this equation the eigenvalue
spectrum can be calculated, for arbitrary values of $x$ and $s$. Because of
the invariance under the substitution $x \rightarrow 1/x$, one can restrict
the ratio $x$ to lie in the interval $0 < x \leq 1$. Similarly, because
of the invariance under the interchange $L_I \leftrightarrow L_{II}$ one
can take $L_{II}$ to be the larger of the two pieces, so that
$s \geq 1$.

In the following we shall impose two simplifying conditions: (i) We take the
tension ratio limit to approach zero,
\beq
x \rightarrow 0.
\label{5}
\eeq
Assuming $T_{II}$ to be a finite quantity, this limit implies that
$T_I \rightarrow 0$.  From the
junction conditions given in \cite{brevik90}
we obtain in this limit the equations
\beq
\xi_I + \e_I = \xi_{II} e^{i\pi\om} + \e_{II} e^{-i\pi\om},
\label{6}
\eeq
\beq
\xi_I e^{2\pi i \om /(1+s)} + \e_I = \xi_{II} e^{2\pi i \om /(1+s)} + \e_{II},
\label{7}
\eeq
\beq
\xi_{II} e^{2\pi i \om} = \e_{II},
\label{8}
\eeq
\beq
\xi_{II} e^{2\pi i \om /(1+s)} = \eta_{II}.
\label{9}
\eeq
According to the dispersion equation (\ref{4}) we obtain now two sequences
of modes.
The eigenfrequencies are seen to be proportional to integers $n$, and will
for clarity be distinguished by separate symbols $\om_n(s)$ and
$\om_n(s^{-1})$:
\beq
\om_n(s) = (1+s)n,
\label{10}
\eeq
\beq
\om_n(s^{-1}) = (1+ s^{-1})n,
\label{11}
\eeq
with $n = \pm 1, \pm 2, \pm 3,...$, corresponding to the first and
the second branch.

(ii) Our second condition is that the length ratio $s$ is an integer,
$s = 1,2,3,\cdots$.

\section{Classical String in Flat $D$-Dimensional Spacetime}\label{III}
\subsection{Oscillator Coordinates. The Hamiltonian}

We are now able to generalize the theory. We consider henceforth the motion
of a two-piece classical string in flat $D$-dimensional space-time.
Following the notation in \cite{green87}
we let $X^\mu(\s,\tau)\,\,\,(\mu = 0,1,2,\cdots
(D-1))$ specify the coordinates on the world sheet. For each of the two
branches - corresponding to Eqs. \ref{10} and \ref{11} respectively -
we can write the general
expression for $X^\mu$ in the form
\beq
X^\mu = x^\mu + \frac{p^\mu\tau}{\pi \bar{T}(s)} + \th(L_I - \s)X_I^\mu
+ \th(\s - L_I)X_{II}^\mu,
\label{12}
\eeq
where $x^\mu$ is the center of mass position and $p^\mu$ is the total
momentum of the string. Besides $\bar{T}(s)$ denotes the mean tension,
\beq
\bar{T}(s) = \frac{1}{\pi}(L_IT_I + L_{II}T_{II})~ \rightarrow~
\frac{s}{1+s}T_{II}.
\label{13}
\eeq
The second term in (\ref{12}) implies that the string's translational
energy $p^0$ is set equal to $\pi \bar{T}(s)$. This generalizes the relation
$p^0 = \pi T$ that is known to hold for a uniform string \cite{green87}.
The two last terms in (\ref{12}) contain the step function,
$\th(x > 0) = 1,\,\,\,\th(x < 0) = 0$. To show the structure of the 
decomposition
of $X^\mu$ into fundamental model we give here the expressions for $X_I^\mu$
for each of the two branches: for the first branch
\beq
X_I^\mu = \frac{i}{2} l(s) \sum_{n \neq 0} \frac{1}{n} \left[
\al_n^\mu(s) e^{i(1+s)n(\s-\tau)} + \tilde{\al}_n^\mu(s)e^{-i(1+s)n(\s+\tau)}
\right],
\label{14}
\eeq
where the $\al_n, \tilde{\al}_n$ are oscillator coordinates of the right-
and left-moving waves respectively. The sum over $n$ goes over all positive and
negative integers except from zero. The factor $l(s)$ is a constant. For the
second branch in region I, analogously
\beq
X_I^\mu=\frac{i}{2} l(s^{-1})\sum_{n \neq 0} \frac{1}{n} \left[
\al_n^\mu(s^{-1}) e^{i(1+s^{-1})n(\s-\tau)}
+ \tilde{\al}_n^\mu(s^{-1})e^{-i(1+s^{-1})n(\s+\tau)} \right],
\label{15}
\eeq
where $l(s^{-1})$ is another constant, which in principle can be different from $l(s)$.
Since $X^\mu$ is real, we must have
\beq
\al_{-n}^\mu = (\al_n^\mu)^*, ~~~~\tilde{\al}_{-n}^\mu =
(\tilde{\al}_n^{\mu})^*.
\label{16}
\eeq
When writing expressions (\ref{14}) and (15), we made use of Eqs.
(\ref{10}) and
(\ref{11}) for the eigenfrequencies. The condition $x \rightarrow 0$
was thus used. The condition that $s$ be an integer has however not so
far been used. This condition will be of importance when we construct
the expression for $X_{II}^\mu$. Before doing this, let us however
consider the constraint equation for the composite string.
Conventionally, when the string is uniform the two-dimensional
energy-momentum tensor $T_{\al \be}\,\,\, (\al,\be = 0,1)$, obtainable as
the variational derivative of the action $S$ with respect to the
two-dimensional metric, is equal to zero. In particular, the
energy density component is then $T_{00} = 0$ locally. In the
present case the situation is more complicated, due to the fact
that the presence of the junctions restricts the freedom of the
variations $\delta X^\mu$. We cannot put $T_{\al \be} = 0$ locally
anymore. What we have at our disposal, is the expression for the
action
\beq
S = -\frac{1}{2}\int d\tau d\s T(\s) \e^{\al \be} \partial_\al X^\mu
\partial_\be X_\mu,
\label{17}
\eeq
where $T(\s)$ is the position-dependent tension
\beq
T(\s) = T_I + (T_{II} - T_I) \th(\s - L_I).
\label{18}
\eeq
The momentum conjugate to $X^\mu$ is $P^\mu(\s) = T(\s) \dot{X}^{\mu}$.
The Hamiltonian of the two-dimensional sheet becomes accordingly
(here $L$ is the Lagrangian)
\beq
H = \int_{0}^{\pi} \left[ P_\mu(\s) \dot{X}^\mu - L \right] d\s =
\frac{1}{2} \int_{0}^{\pi} T(\s) (\dot{X}^2 + {X'}^2)d\s.
\label{19}
\eeq
The basic condition that we shall impose, is that $H = 0$ when
applied to the physical states. This is a more weak condition than
the strong condition $T_{\al \be} = 0$ applicable for a uniform string.

\subsection{Classical Mass Formula. The First Branch}

Assume  that $s$ is an arbitrary integer,  $s = 1,2,3,\cdots$. When $s$ is different from 1, we have to
distinguish between the eigenfrequencies $\om_n(s)$ and $\om_n(s^{-1})$ for
the first and the second branch.
Let us consider the first  branch. In region I, the representation for the right- and left-
moving modes was given above, in Eq. (\ref{14}). For reasons that will
become clear from the quantum mechanical discussion later, we will choose
$l(s)$ equal to
\beq
l(s) = \frac{1}{\sqrt{\pi T_I}}\,\,.
\label{20}
\eeq
Since we have assumed $T_I$ to be small, the expression (20) will tend to 
infinity. This is actually  a delicate point: as we will see later, in 
Eq. (\ref{31}), the
expression for the Hamiltonian $H_I$ in region I derived with the
use of Eq. (\ref{20}) is {\it independent}
of the tension. Thus $H_I$ behaves in a 'normal' way.
The coordinates $X_I^\mu$ in region I themselves diverge, but after
multiplication with the very small string tension  $T_I$ they become 
suppressed and lead to a finite expression for $H_I$.  Accordingly, if one 
quantizes the system starting from the canonical commutaion relations in 
region I ,  one arrives at a standard quantum mechanical
picture in this region. (We will return to this point in Sec. \ref{IV}.)

When writing the analogous mode expansion in region II, we have to observe the
junction conditions (\ref{6}) - (\ref{9}), which hold for all $s$. For the first
branch $\om_n(s)$, and for {\it odd} values of $s$, it is seen that the junction conditions impose no
restriction on the values of $n$. {\it All} frequencies, corresponding to
 $n = \pm1,\pm2,\pm3,\cdots$, permit the waves to propagate
from region I to region II. Equations (\ref{6}) - (\ref{9}) reduce in
this case to the equations
\beq
\xi_{I} + \e_{I} = 2\xi_{II} = 2\e_{II},
\label{21}
\eeq
which show that the right- and left-moving amplitudes $\xi_{I}$ and
$\e_{I}$ in region I can be chosen freely and that the amplitudes
$\xi_{II},\e_{II}$ in region II are thereafter fixed. Transformed into
oscillator coordinate language, this means that $\al_n^\mu$ and
$\tilde{\al}_n^\mu$ can be chosen freely.

If $s$ is an {\it even} integer, then the validity of Eqs.(21) requires $n$
in Eq.(10) to be even. If $n$ is odd, the junction conditions reduce instead
to
\beq
\xi_I + \eta_I = 0, \,\,\,   \xi_{II} = \eta_{II} = 0,
\label{22}
\eeq
which show that the waves are now unable to penetrate into region II. The oscillations in
 region I are in this case standing waves.

The expansion for the first branch in region II can in view of (\ref{21})
be written
\beq
X_{II}^\mu = \frac{i}{2\sqrt{\pi T_I}} \sum_{n \neq 0} \frac{1}{n}
\g_n^\mu(s)e^{-i(1+s)n \tau} \cos[(1+s)n \s],
\label{23}
\eeq
where we have defined $\g_n(s)$ as
\beq
\g_n^\mu(s) = \al_n^\mu(s) + \tilde{\al}_n^\mu(s), ~~~~n \neq 0.
\label{24}
\eeq
The oscillations in region II are thus standing waves; this being a direct
consequence of the condition $x \rightarrow 0$.

It is useful to introduce light-cone coordinates, $\s^- = \tau - \s$ and
$\s^+ = \tau + \s$. The derivatives conjugate to $\s^\mp$ are
$\partial_\mp = \frac{1}{2}(\partial_\tau \mp \partial_\s)$. In region I we
find
\begin{eqnarray}
\partial_{-}X^\mu &=& \frac{1+s}{2\sqrt{\pi T_I}} \sum_{-\infty}^{\infty}
\alpha_{n}^\mu(s)e^{i(1+s)n(\sigma-\tau)},\nonumber\\
\partial_{+}X^\mu &=& \frac{1+s}{2\sqrt{\pi T_I}} \sum_{-\infty}^{\infty}
\tilde{\alpha_{n}}^\mu(s)e^{-i(1+s)n(\sigma+\tau)},
\label{25}
\end{eqnarray}
where we have defined
\beq
\al_0^{\mu}(s) = \tilde{\al}_0^\mu(s) = \frac{p^{\mu}}{T_{II}s}
\sqrt{\frac{T_I}{\pi}}
\mbox{.}
\label{26}
\eeq

Further, in region II  we find
\beq
\partial_{\mp} X^\mu = \frac{1+s}{4 \sqrt{\pi T_I}}
\sum_{-\infty}^{\infty}
 \g_n^\mu(s)
e^{\pm i(1+s) n (\s \mp \tau)},
\label{27}
\eeq
with
\beq
\g_0^\mu(s) = \frac{2p^\mu}{T_{II}s} \sqrt{\frac{T_I}{\pi}} =
2 \al_0^\mu(s).
\label{28}
\eeq

Inserting Eqs. (\ref{24}) and (\ref{27}) into the Hamiltonian
\beq
H = \int_0^{\pi} T(\s)(\partial_{-}X \cdot \partial_{-}X + \partial_{+}X \cdot
\partial_{+}X) d\s
\label{29}
\eeq
we get, for the full first branch
\beq
H = H_I + H_{II}
\mbox{,}
\label{30}
\eeq
where
\begin{eqnarray}
H_I & = & T_I \int_{I} (\partial_-X \cdot \partial_-X + \partial_+X
\cdot \partial_+X) d\s\cr
  & &\cr
    & = & \frac{1+s}{4} \sum_{-\infty}^{\infty} [\al_{-n}(s) \cdot \al_n(s)
+ \tilde{\al}_{-n}(s) \cdot \tilde{\al}_n(s)],
\label{31}
\end{eqnarray}
\begin{eqnarray}
H_{II} & = & T_{II} \int_{II} (\partial_-X \cdot \partial_-X + \partial_+X
\cdot \partial_+X) d\s\cr
   & &\cr
    & = & \frac{s(1+s)}{8x} \sum_{-\infty}^{\infty} \gamma_{-n}(s) \cdot
\gamma_n(s),
\label{32}
\end{eqnarray}
with $x = T_I/T_{II}$ as before.

The case $s = 1$ is of particular interest. The string is then divided into
two pieces of equal length. We have then
\beq
H_I(s=1) = \frac{1}{2}\sum_{-\infty}^{\infty} \left(
\al_{-n} \cdot \al_n + \tilde{\al}_{-n} \cdot \tilde{\al}_n \right),
\label{33}
\eeq
\beq
H_{II}(s=1) = \frac{1}{4x} \sum_{-\infty}^{\infty} \g_{-n} \cdot \g_n.
\label{34}
\eeq
It is notable that Eq. (33) is formally the same as the standard
expression for a closed uniform string, of length $\pi$. See, for instance,
Eq. (2.1.76) in Ref. \cite{green87}. The fact that we recover the
characteristics of a closed string in region I is understandable, since
this part of our composite string permits both right-moving and
left-moving waves. The absence of any tension-dependent factor in front of
the expression (\ref{33}) is related to our choice (\ref{20}) for the
length $l(s)$. Moreover, Eq. (\ref{34}) is essentially the standard expression
for en {\it open} uniform string, corresponding to standing waves. The
presence of the inverse tension ratio $1/x$ in front of the expression is
caused by the junction conditions, Eqs. (\ref{21}).

The condition $H=0$ enables us to calculate the mass $M$ of the string. It
must be given by $M^2 = -p^\mu p_\mu$, similarly as in the uniform string case
\cite{green87}. From Eqs. (\ref{31}) and (\ref{32}) we obtain, taking into
account that $x << 1$ and that
$\al_0(s) \cdot \al_0(s) =-M^2x/(\pi T_{II} s^2)$,
\beq
M^2 = \pi T_{II} s  \sum_{n=1}^{\infty} \left[ \al_{-n}(s) \cdot
\al_n(s) + \tilde{\al}_{-n}(s) \cdot \tilde\al_n(s) +
\frac{s}{2x} \g_{-n}(s) \cdot \g_n(s) \right].
\label{35}
\eeq
This holds for the first branch, for odd/even values of $s$.

\subsection{The Second Branch}

For the second branch whose eigenfrequencies are $\om (s^{-1})$ the mode 
expansion in region I becomes 

\beq
X_I^\mu = \frac{i}{2\sqrt{\pi T_I}}\sum_{n \neq 0} \frac{1}{n}
\left[
\al_{n}^\mu(s^{-1})e^{i(1+s^{-1})n(\s-\tau)} +
\tilde{\al}_{n}^\mu(s^{-1})e^{-i(1+s^{-1})n(\s+\tau)}
\right].
\label{36}
\eeq

Analogously in region II

\beq
X_{II}^\mu = \frac{i}{2\sqrt{\pi T_I}} \sum_{n \neq 0} \frac{1}{n}
\g_{n}^\mu(s^{-1})e^{-i(1+s^{-1})n\tau} \cos (1+s^{-1})n\s,
\label{37}
\eeq

where
\beq
\g_{n}^\mu(s^{-1}) = \al_{n}^\mu(s^{-1}) + \tilde{\al}_{n}^\mu(s^{-1}),~~ 
n\neq 0
\mbox{.}
\label{38}
\eeq
The expansions (36) and (37) hold for all integers $s$. This is so because
the basic
expressions (10) and (11) for the eigenfrequencies hold for all values of $s$.
However it may be noted that if the junction conditions are required to imply
nonvanishing oscillations in region II, corresponding to nonvanishing right
hand sides
in Eq.(21), then further restrictions come into play. Namely, if $s$ is odd,
the index $n$ in Eqs.(36) and (37) has to be a multiple of $s$.
If $s$ is even, then $n$ has to be an {\it even} integer times $s$.
We recall that analogous considerations were made in the case of the first
branch.
When we later shall consider the quantum mechanical free energy, it becomes
necessary
to include {\it all} modes, including those that lead to zero oscillations in
region II
according to the classical theory.

Let us calculate the light-cone derivatives: in region I they are

\begin{eqnarray}
\partial_{-}X^\mu &=& \frac{1+s^{-1}}{2\sqrt{\pi T_I}} \sum_{-\infty}^{\infty}
\al_{n}^{\mu}(s^{-1})e^{i(1+s^{-1})n(\s-\tau)},\nonumber\\
\partial_{+}X^\mu & = & \frac{1+s^{-1}}{2\sqrt{\pi T_I}} \sum_{-\infty}^{\infty}
\tilde{\al_{n}^{\mu}}(s^{-1})e^{-i(1+s^{-1})n(\s+\tau)},
\label{39}
\end{eqnarray}
and in region $II$
\beq
\partial_{\mp}X^\mu = \frac{1+s^{-1}}{4\sqrt{\pi T_I}} \sum_{-\infty}^{\infty}
\g_{n}^\mu(s^{-1})e^{\pm i(1+s^{-1})n(\s \mp \tau)},
\label{40}
\eeq
where
\beq
\al_0^\mu(s^{-1}) = \tilde{\al_0^\mu}(s^{-1})= \frac{1}{2} \g_0^\mu(s^{-1})
= \frac{p^\mu}{T_{II}} \sqrt{\frac{T_I}{\pi }}
\mbox{.}
\label{41}
\eeq
Thus $\al_0(s^{-1})$ differs from $\al_0(s)$, Eq.(26).
Again writing the Hamiltonian as $H = H_I + H_{II}$, we now find
\beq
H_I = \frac{1+s^{-1}}{4s} \sum_{-\infty}^{\infty}
\left[
\al_{-n}(s^{-1}) \cdot \al_{n}(s^{-1}) + \tilde{\al}_{-n}(s^{-1}) \cdot
\tilde{\al}_{n}(s^{-1})
\right],
\label{42}
\eeq
\beq
H_{II} = \frac{1+s^{-1}}{8x} \sum_{-\infty}^{\infty}
\g_{-n}(s^{-1}) \cdot \g_{n}(s^{-1}).
\label{43}
\eeq
If $s=1$, we recover the same expressions for $H_I$ and $H_{II}$, Eqs.
(\ref{33}) and (\ref{34}), as for the first branch.

>From the condition $H=0$ we calculate the mass, observing that
$\al_0(s^{-1}) \cdot \al_0(s^{-1}) = -M^2x/(\pi T_{II}$):
\begin{eqnarray}
M^2 & = & \frac{\pi T_{II}}{s} \sum_{n=1}^{\infty}
\left[
\al_{-n}(s^{-1}) \cdot \al_{n}(s^{-1}) + \tilde{\al}_{-n}(s^{-1})\cdot
\tilde{\al}_{n}(s^{-1})\right] \nonumber\\
& + & \frac{\pi T_{II}}{2x}\sum_{n=1}^{\infty}\g_{-n}(s^{-1})\cdot\g_{n}(s^{-1}).
\label{44}
\end{eqnarray}

\section{Quantum Theory. The Free Energy of the String}\label{IV}
\subsection{Quantization}

We shall consider the free energy of the quantum fields with masses given
by the mass formula corresponding to the piecewice bosonic string. 
Our starting point is the following expression for the free energy $F$, at 
finite temperature $T$, of free fields of mass $M$ in $D$-dimensions: 
\beq
\be F = -\ln Z = \frac{1}{2} \be \sum_{-\infty}^{\infty} \om_n - \be
\sum_{m=1}^{\infty} \int_{0}^{\infty} \frac{du}{u} (2 \pi u)^{-D/2}
\exp \left( - \frac{M^2u}{2} - \frac{m^2\be^2}{2u} \right).
\label{45}
\eeq
Here $\be = 1/k_BT$, and $Z$ is the partition function. We follow the formalism
of Ref. \cite{alvarez87}; some other related references are
\cite{sundborg85} - \cite{bytsenko96}. The constituent ``fields'' of the 
quantum gas are the excitations associated with the modes of a single string.

In Eq.(\ref{45}) we thus need to know the expression for $M^2$ in the
quantum theory. We quantize the system according to conventional methods as
found, for instance, in Ref. \cite{green87},
Chapter 2.2. In accordance with the canonical prescription in region I the 
equal-time commutation rules are required to be

\beq
T_{I}[\dot{X}^\mu(\s,\tau), X^\nu(\s', \tau)]=
-i \delta(\s-\s')\e^{\mu \nu},
\label{46}
\eeq
and in region II
\beq
T_{II} [\dot{X}^\mu(\s,\tau),X^\nu(\s',\tau)] =
-i \delta(\s - \s')\e^{\mu \nu},
\label{47}
\eeq
where $\e^{\mu \nu}$ is the $D$-dimensional metric. These relations are in
conformity with the fact that the momentum conjugate to $X^\mu$ is in either
region equal to $T(\s) \dot{X}^\mu$. The remaining commutation relations
vanish:
\beq
[X^\mu(\s,\tau),X^\nu(\s',\tau)] = [\dot{X}^\mu(\s,\tau),\dot{X}^\nu(\s',\tau)]
 = 0.
\label{48}
\eeq

The quantities to be promoted to Fock state operators are
$\al_{\mp n}(s)$ and $\tilde{\al}_{\mp n}(s)$ (first branch, region I),
$\g_{\mp n}(s)$ (first branch, region II),
$\al_{\mp n}(s^{-1})$ and $\tilde{\al}_{\mp n}(s^{-1})$ (second branch,
region I), and $\g_{\mp n}(s^{-1})$ (second branch, region II). These
operators satisfy
\begin{eqnarray}
\al_{-n}^{\mu}(s) &=& \al_{n}^{\mu\dagger}(s),\,\,\,
\g_{-n}^{\mu}(s) = \g_{n}^{\mu\dagger}(s), \nonumber\\
\al_{-n}^{\mu}(s^{-1}) &=& \al_{n}^{\mu\dagger}(s^{-1}),\,\,\,
\g_{-n}^{\mu}(s^{-1}) = \g_{n}^{\mu\dagger}(s^{-1})
\label{49}
\end{eqnarray}
for all $n$.
We insert our previous expansions for  $X^\mu$ and $\dot{X}^\mu$ in the
commutation relations in regions I and II for the two branches, and make use
of the effective relationship
\beq
\sum_{-\infty}^{\infty}e^{i(1+s)n(\s-\s')}=
2\sum_{-\infty}^{\infty}\cos(1+s)n\s\cos(1+s)n\s'
\rightarrow \frac{2\pi}{1+s}\delta(\s-\s').
\label{50}
\eeq
For the first branch we then get in region I

\beq
[\al_n^\mu(s),\al_m^\nu(s)] = n \de_{n+m,0} \e^{\mu\nu},
\label{51}
\eeq
with a similar relation for the $\tilde{\al}_n$. In region II

\beq
[\g_{n}^\mu(s),\g_{m}^\nu(s)] = 4nx \de_{n+m,0} \e^{\mu\nu}.
\label{52}
\eeq
For the second branch we get analogously
\beq
[\al_{n}^{\mu}(s^{-1}),\al_{m}^{\nu}(s^{-1})]=n\de_{n+m,0}\e^{\mu\nu},
\,\,\,\,\,
[\g_{n}^{\mu}(s^{-1}), \g_{m}^{\nu}(s^{-1})]= 4nx\de_{n+m,0}\e^{\mu\nu}.
\label{53}
\eeq
By introducing annihilation and creation operators for the first branch in the
following way:
\begin{eqnarray}
\al_n^\mu(s) & = & \sqrt{n} a_n^\mu(s),\,\,\,\,\,
\al_{-n}^\mu(s)=\sqrt{n} a_n^{\mu\dagger}(s), \nonumber\\
\g_n^\mu(s) &=& \sqrt{4nx} c_{n}^{mu}(s),\,\,\,\,\,
\g_{-n}^\mu(s) =
\sqrt{4nx} c_{n}^{\mu\dagger}(s),
\label{54}
\end{eqnarray}
we find for $n \geq 1$ the standard form
\beq
[a_{n}^{\mu}(s), a_{m}^{\nu\dagger}(s)] = \de_{nm}\e^{\mu\nu},
\label{55}
\eeq

\beq
[c_{n}^{\mu}(s), c_{m}^{\nu\dagger}(s)] = \de_{nm}\e^{\mu\nu}.
\label{56}
\eeq
The commutation relations for the second branch are analogous,
only with the replacement $s\rightarrow s^{-1}$.

\subsection{The Free Energy and the Hagedorn Temperature}

In the following we shall limit ourselves to the first branch only. Using Eqs.
(54) in
Eqs.(31) and (32) we may write the two parts of the Hamiltonian as

\beq
H_{I} = -\frac{M^2x}{2st(s)} + \frac{1}{2}\sum_{n=1}^{\infty} \om_n(s)
[a_n^{\dagger}(s)\cdot a_n(s) + \tilde{a_n}^{\dagger}(s)\cdot\tilde{a_n}(s)],
\label{57}
\eeq
\beq
H_{II} = -\frac{M^2}{2t(s)} + s \sum_{n=1}^{\infty} \om_n(s)
c_n^{\dagger}(s) \cdot c_n(s),
\label{58}
\eeq
where we for convenience have introduced the symbol $t(s)$ defined by
\beq
t(s) = \pi\bar{T}(s).
\label{59}
\eeq
(Observe the notation  $c_n^{\dagger}\cdot c_n \equiv c_n^{\mu\dagger}
c_{n\mu}$). The extra factor $s$ in Eq. (\ref{58}) is
related to the fact that the relative length of region II is equal to
$s$.
>From the condition $H = H_{I}+H_{II} = 0$ in the limit $x \rightarrow 0$
we obtain, either from Eqs. (57) and (58) or directly from Eq. (35),
\begin{eqnarray}
M^{2} &=& t(s)\sum_{i=1}^{24}\sum_{n=1}^{\infty}\om_n(s)
[a_{ni}^{\dagger}(s)a_{ni}(s) + \tilde{a}_{ni}^{\dagger}\tilde{a}_{ni}(s)
- A_1] \nonumber\\
&+& 2st(s)\sum_{i=1}^{24}\sum_{n=1}^{\infty}\om_n(s)[c_{ni}^{\dagger}(s)c_{ni}(s)
- A_2].
\label{60}
\end{eqnarray}
We have here put $D = 26$, the commonly accepted space-time dimension for the
bosonic string.
As usual, the $c_{ni}$ denote the transverse oscillator operators  (here for
the first branch). Further, we have introduced in Eq.(60) two constants
$A_1$ and $A_2$, in order to take care of ordering ambiguities.

In Eq. (\ref{45}), there occurs a zero-point energy $\frac{1}{2} \sum \om_n$,
summed over all eigenfrequencies. Apart from an infinite constant of no
physical significance, this is actually the Casimir energy, which was
calculated in \cite{brevik90}. When $x \rightarrow 0$ we have, for
arbitrary $s$, when the string length equals $\pi$,
\beq
\frac{1}{2} \sum_{- \infty}^{\infty} \om_n \rightarrow
- \frac{1}{24}(s + \frac{1}{s} - 2).
\label{61}
\eeq

The constraint for the closed string (expressing the invariance of the theory
in the region I under shifts of the origin of the co-ordinate) has the form
\beq
\sum_{i=1}^{24}\sum_{n=1}^{\infty}\om_n(s)\left[
a_{ni}^{\dagger}(s)a_{ni}(s) - \tilde{a}_{ni}^{\dagger}\tilde{a}_{ni}(s)
\right]=0
\mbox{.}
\label{62}
\eeq
The commutation relations for above operators are given by Eqs.(55) and (56).
The mass of state (obtained by acting on the Fock vacuum $|0>$ with creation
operators) can be written as follows
$({\rm mass})^2\sim a_{n1}^{\dagger}...a_{ni}^{\dagger}c_{n1}^{\dagger}...
c_{ni}^{\dagger}|0>$. As usual the physical Hilbert space consists of all Fock
space states obeying the condition (62), which can be implemented by means
of the integral representation for Kronecker deltas. Thus the free energy
of the field content in the "proper time" representation becomes

\begin{eqnarray}
&F&=-\frac{1}{24}(s+\frac{1}{s}-2) \nonumber\\
&-&2^{-14}\pi^{-13}\int_0^{\infty}\frac{d\tau_2}{\tau_2^{14}}
\left[\theta_3\left(0|\frac{i\beta^2}{2\pi\tau_2}\right)-1\right]
{\rm Tr}\exp\left\{-\frac{\tau_2M^2}{2}\right\} \nonumber\\
& \times & \int_{-\pi}^{\pi}\frac{d\tau_1}{2\pi}{\rm Tr}
\exp\left\{i\tau_1\sum_{i=1}^{24}\sum_{n=1}^{\infty}\om_n(s)\left[
a_{ni}^{\dagger}(s)a_{ni}(s) - \tilde{a}_{ni}^{\dagger}(s)\tilde{a}_{ni}(s)
\right]\right\}.
\label{63}
\end{eqnarray}
Performing the trace over the entire Fock space (note that $[H_I,H_{II}]=0$
and ${\rm Tr}\,y^{a_n^{\dagger}a_n}=(1-y)^{-1}$) we have
\begin{eqnarray}
{\rm Tr}\exp\left\{\sum_{i=1}^{24}\sum_{n=1}^{\infty}\om_n(s)
a_{ni}^{\dagger}(s)a_{ni}(s)\left(-\frac{1}{2}t(s)\tau_2\pm i\tau_1\right)
\right\}\nonumber\\
=\prod_{n=1}^{\infty}\left[1-e^{\om_n(s)(-\frac{1}{2}t(s)\tau_2\pm i\tau_1)}
\right]^{-24},
\label{64}
\end{eqnarray}
\begin{eqnarray}
{\rm Tr}\exp\left\{-st(s)\tau_2\sum_{i=1}^{24}\sum_{n=1}^{\infty}\om_n(s)
c_{ni}^{\dagger}(s)c_{ni}(s)\right\}\nonumber\\
=\prod_{n=1}^{\infty}\left[1-e^{-st(s)\tau_2\om_n(s)}\right]^{-24}.
\label{65}
\end{eqnarray}
Working out the sums in Eq. (63) for $A_1=2,\,\,A_2=1$, and changing 
variables to $\tau_1\rightarrow\tau_1 2\pi,\,\tau_2\rightarrow 
\tau_2 4\pi/t(s)$ one can finally get
\begin{eqnarray}
F&=&-\frac{1}{24}(s+\frac{1}{s}-2) \nonumber\\
&&-2^{-40}\pi^{-26}t(s)^{-13}\int_0^{\infty}\frac{d\tau_2}{\tau_2^{14}}
\int_{-1/2}^{1/2}d\tau_1\left[\theta_3\left(0|\frac{i\beta^2t(s)}
{8\pi^2\tau_2}\right)-1\right] \nonumber\\
&& \times|\eta[(1+s)\tau]|^{-48}\eta[s(1+s)(\tau-\overline{\tau})]^{-24},
\label{66}
\end{eqnarray}
where we integrate over all possible non-diffeomorphic toruses which are
characterized by a single Teichm{\"u}ller parameter $\tau=\tau_1+i\tau_2$.
In Eq. (66) the Dedekind $\eta$-function and the Jacobi $\theta_3$-function
\beq
\e(\tau) = e^{\frac{\pi i\tau}{12}}\prod_{n=1}^{\infty}(1 - e^{2\pi in\tau}),
\label{67}
\eeq
\beq
\th_3(v|x) = \sum_{n=-\infty}^{\infty} e^{ixn^2+2 \pi ivn},
\label{68}
\eeq
and the condition $\eta(-\overline{\tau})=\overline{\eta(\tau)}$ has been
used.

Once the free energy has been found, the other thermodynamic quantities can
readily be calculated. For instance, the energy $U$ and the entropy $S$ of
the system are
\beq
U = \frac{\partial (\be F)}{\partial \be},
~~~~~~~~ S = k_B \be^2 \frac{\partial F}{\partial \be}.
\label{69}
\eeq

What is the Hagedorn temperature, $T_c = 1/(k_B \be_c)$, of the composite
string? This critical temperature, introduced by Hagedorn in the context
of strong interactions a long time ago \cite{hagedorn65}, is the
temperature above which the free energy is ultraviolet divergent. In the
ultraviolet limit ($\tau_2 \rightarrow 0$),
\begin{eqnarray}
\e^{-24} (i \tau) &=& \tau^{12} e^{2 \pi/ \tau}\left[1+O\left
(e^{-2\pi/\tau}\right)\right],\nonumber\\
\th_3 \left( 0|\frac{i\be^2t(s)}{8\pi^2\tau_2} \right)-1
& = &
2 \exp \left( - \frac{\beta^2t(s)}{8\pi^2\tau_2}\right)+
O\left(\exp\left(-\frac{\beta^2t(s)}{2\pi^2\tau_2}\right)\right),
\label{70}
\end{eqnarray}
which upon insertion into Eq. (66) shows that the integrand is ultraviolet
finite if
\beq
\be > \be_c = \frac{4}{s}\sqrt{\frac{\pi(1+s)}{T_{II}}}.
\label{71}
\eeq

For a fixed value of $T_{II}$ the Hagedorn temperature is thus seen to depend
on $s$. We may mention here that the physical meaning of the Hagedorn
temperature is still not clear. There are different interpretations
possible: {\it (i)} one may argue that $T_c$ is the maximum obtainable
temperature in string systems, this meaning, when applied to cosmology,
that there is a maximum temperature in the early Universe. Or, {\it (ii)} one
may take $T_c$ to indicate some sort of phase transition to a new stringy
phase. Some further discussion on these matters is given, for instance,
in Refs. \cite{alvarez87,alvarez89,elizalde94,bytsenko96}.

Finally, let us consider the limiting case in which one of the pieces of the
string is much shorter than the other. Physically this case is of interest,
since it corresponds to a point mass sitting on a string (note, however,
that the "mass" concept must be taken in a peculiar sense, since the
relativistic condition of Eq. (\ref{1}) is satisfied always). Since we have
assumed that $s \geq 1$, this case corresponds to $s \rightarrow\infty$. We
let the tension $T_{II}$ be fixed, though arbitrary.
It is seen, first of all, that the Hagedorn temperature (\ref{71}) goes to
infinity so that $F$ is always ultraviolet finite,
\beq
\be_c \rightarrow 0, ~~~~~~~~ T_c \rightarrow \infty.
\label{72}
\eeq
Next, since $\exp\left(-\beta^2t(s)/8\pi^2\tau_2\right)$ can be
taken to be small we obtain, when using again the expansion (\ref{70}) for
$\th_3\left(0|i\beta^2t(s)/8\pi^2\tau_2\right)$,
\begin{eqnarray}
F_{(\beta\rightarrow 0)} &=& -\frac{s}{24}-
(8\pi^3T_{II})^{-13}\int_0^{\infty}\frac{d\tau_2}{\tau_2^{14}}
\int_{-1/2}^{1/2}d\tau_1 \nonumber\\
& \times & \exp\left(-\frac{\beta^2T_{II}}{8\pi\tau_2}\right)
|\eta[(1+s)\tau]|^{-48}\eta[s(1+s)(\tau-\overline{\tau})]^{-24}.
\label{73}
\end{eqnarray}
Physically speaking, the linear dependence of the first term in (\ref{73})
reflects that the Casimir energy of a little piece of string
embedded in an essentially infinite string has for dimensional reasons to be
inversely proportional to the length $L_I = \pi/(1+s) \simeq \pi/s$ of
the little string. The first term in (\ref{73}) is seen to outweigh the
second, integral term, which goes to zero when $s \rightarrow \infty$.

\newpage
\section{Summary}

For the two-piece relativistic string, the starting point in classical
theory is the dispersion equation (\ref{4}), valid for arbitrary tension
ratios $x = T_I/T_{II}$ and length ratios $s = L_{II}/L_I$. In our
calculations we have made two simplifying assumptions: first, we have
considered only the limit $x \rightarrow 0$. Taking $T_{II}$ to be
finite, this means that $T \rightarrow 0$. Second, we have assumed $s$
to be an integer.

The string's eigenvalue spectrum is given by Eqs. (\ref{10}) and (\ref{11}),
meaning that there are in general (for $s \neq 1$) two different branches.
The boundary conditions at the two junctions are given by Eqs. (\ref{6}) -
(\ref{9}). As for the first branch, Eq. (\ref{10}), there is for odd $s$ no 
restriction on
$n$. All frequencies are permitted to propagate from region I to region II.
The junction conditions reduce to one single equation,  Eq. (\ref{21}). If $s$
is even, the waves are unable to propagate into region II. The oscillations in
region I are then standing waves.
As for the second branch, Eq.(11), and for odd values of $s$,  the integer $n$
in Eqs.(36) and (37) has to be a multiple of $s$ in order to permit 
nonvanishing oscillations
in region II.  If $s$ is even, then $n$ has correspondingly to be an even 
integer times $s$.

In some sense the behaviour of the string in region II is similar to that of 
a conventional open
uniform string of length $L_{II}$. The physical conditions in the two cases 
are however not
the same, since Eqs. (\ref{6}) - (\ref{9})  are different from the
boundary conditions
$X^{'\mu} = 0$ at the ends of an open string \cite{green87}. Our
 string model is actually some kind of a hybrid model.
In the quantum mechanical formulation the starting point for the free energy $F$
is Eq.(45), into which we have to insert the expression (62) for the quadratic 
mass. The final result for $F$ is given by Eq.(66). Once $F$ is known, other
thermodynamic quantities can easily be calculated. The inverse Hagedorn
temperature is given by Eq.(71), and is dependent on $s$. If
$s \rightarrow \infty$, corresponding to a point "mass" sitting on
the string, the Hagedorn temperature goes to infinity.

\section{Acknowledgment}

I. Brevik thanks NORDITA for financial support to a visit to Copenhagen.
A.A. Bytsenko wishes to thank Nordic Scholarship and the Department
of Applied Mechanics of Norwegian University of Science and Technology
(Trondheim) for financial support and kind hospitality. The research of A.A.
Bytsenko was supported in part by the CNPq grant, the Russian Foundation for
Basic Research (grant No. 98-02-18380-a) and by GRACENAS (grant
No. 6-18-1997).


\newpage
\begin{thebibliography}{99}

\bibitem{brevik90}
I. Brevik and H.B. Nielsen,
Phys. Rev. D {\bf 41}, 1185 (1990).

\bibitem{li91}
X. Li, X. Shi and J. Zhang,
Phys. Rev. D {\bf 44}, 560 (1991).

\bibitem{brevik94}
I. Brevik and E. Elizalde,
Phys. Rev. D {\bf 49}, 5319 (1994).
Cf. also E. Elizalde, {\it Ten Physical
Applications of Spectral Zeta Functions}, (Springer, Berlin, 1995), Sec. 7.2.

\bibitem{brevik95}
I. Brevik and H.B. Nielsen,
Phys. Rev. D {\bf 51}, 1869 (1995).

\bibitem{brevik96}
I. Brevik, H.B. Nielsen and S.D. Odintsov,
Phys. Rev. D {\bf 53}, 3224 (1996).

\bibitem{bayin96}
S.S. Bayin, J.P. Krisch and M. Ozcan,
J. Math. Phys. {\bf 37}, 3662 (1996).

\bibitem{brevik}
I. Brevik and R. Sollie, J. Math. Phys. {\bf 38}, 2774 (1997).

\bibitem{berntsen97}
M.H. Berntsen, I. Brevik and S.D. Odintsov, Ann. Phys. (NY) {\bf 257}, 84 
(1997).

\bibitem{elizalde96}
E. Elizalde, M. Kawamura, A. Sugamoto, S. Nojiri and S.D. Odintsov,
Int. J. Mod. Phys. A {\bf 11}, 5569 (1996).

\bibitem{green87}
M.B. Green, J.H. Schwarz and E. Witten,
{\it Superstring Theory}, Vol. {\bf 1} (Cambridge Univ. Press, Cambridge, 
1987).

\bibitem{alvarez87}
E. Alvarez and M.A.R. Osorio,
Phys. Rev. D {\bf 36}, 1175 (1987).

\bibitem{sundborg85}
B. Sundborg, 
Nucl. Phys. B {\bf 254}, 583 (1985). 

\bibitem{olesen86}
P. Olesen,
Nucl. Phys. B {\bf 267}, 539 (1986).

\bibitem{okada87}
H. Okada,
Prog. Theor. Phys. {\bf 77}, 751 (1987).

\bibitem{salomonson89}
P. Salomonson and B.-S. Skagerstam,
Physica A {\bf 158}, 499 (1989).

\bibitem{alvarez89}
E. Alvarez and M.A.R. Osorio,
Physica A {\bf 158}, 449 (1989).

\bibitem{elizalde94}
E. Elizalde, S.D. Odintsov, A. Romeo, A.A. Bytsenko and S. Zerbini,
{\it Zeta Regularization with Applications} (World Scientific, Singapore,
1994), Chapter 8.

\bibitem{bytsenko96}
A.A. Bytsenko, G. Cognola, L. Vanzo and S. Zerbini,
Phys. Reports {\bf 266}, 1 (1996).

\bibitem{hagedorn65}
R. Hagedorn,
Suppl. al Nuovo Cimento {\bf 3}, 147 (1965).
\end{thebibliography}
\end{document}